# CORESET OF HYPERSPECTRAL IMAGES ON A SMALL QUANTUM COMPUTER

*Soronzonbold Otgonbaatar, Mihai Datcu, Begüm Demir*

DLR Oberpfaffenhofen, Technical University of Berlin

## ABSTRACT

Machine Learning (ML) techniques are employed to analyze and process big Remote Sensing (RS) data, and one well-known ML technique is a Support Vector Machine (SVM). An SVM is a quadratic programming (QP) problem, and a D-Wave quantum annealer (D-Wave QA) promises to solve this QP problem more efficiently than a conventional computer. However, the D-Wave QA cannot solve directly the SVM due to its very few input qubits. Hence, we use a coreset ("core of a dataset") of given EO data for training an SVM on this small D-Wave QA. The coreset is a small, representative weighted subset of an original dataset, and any training models generate competitive classes by using the coreset in contrast to by using its original dataset. We measured the closeness between an original dataset and its coreset by employing a Kullback-Leibler (KL) divergence measure. Moreover, we trained the SVM on the coreset data by using both a D-Wave QA and a conventional method. We conclude that the coreset characterizes the original dataset with very small KL divergence measure. In addition, we present our KL divergence results for demonstrating the closeness between our original data and its coreset. As practical RS data, we use *Hyperspectral Image* (HSI) of Indian Pine, USA.

## 1. INTRODUCTION

Most remotely sensed images are massive and diverse to classify by using ML techniques when compared with conventional images. In principle, ML is a machinery for identifying and classifying common patterns in a labeled as well as unlabeled dataset [1]. However, training models are computationally expensive and intractable on big data. Several studies proposed to use only a coreset ("core of a dataset") of an original dataset for tackling these computational expensive and intractability challenges via Bayesian inference [2], [3], [4]. This coreset is a small, representative weighted subset of an original dataset, and any learning models trained on the coreset generate classes competitive with the ones trained on the original dataset (see Fig. 1).

The concept of a coreset creates a new opportunity for training ML models by using small quantum computers [5],

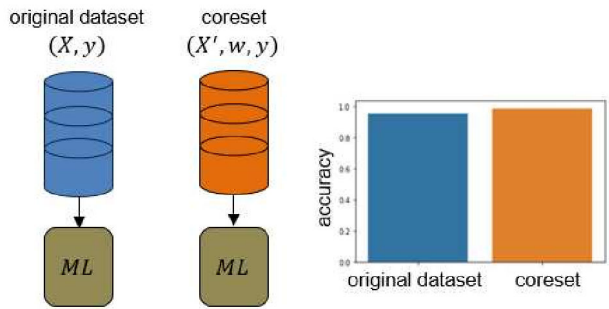

**Fig. 1**. The same learning models (ML) trained on an original dataset $(X, y)$ and its coreset $(X', w, y)$ generate approximately similar results.

[6] since a currently available quantum computer (D-Wave QA) comprises a very few quantum bits (qubits) [7]. Moreover, the quantum computer promises to solve an SVM faster than a conventional method and some intractable problems in ML [8], [9], [10]. We termed a quantum SVM (qSVM) when solving the SVM on a quantum computer, otherwise a classical SVM (cSVM).

For analysing practical RS datasets, there are several approaches of training ML models and solving RS optimization problems on a D-Wave QA [11], [12], [13], and even on a gate-based quantum computer [14], [15]. The D-Wave QA has around $5,000$ input qubits and a specific *Pegasus* topology for the connectivity of its qubits [16], and it is designed for solving a Quadratic Unconstrained Binary Optimization (QUBO) problem [9]. In addition, If we assume that a single variable of a QUBO problem is represented by 6 qubits then we have approximately 833 qubits, and even this number decreases due to its topology. Therefore, in this work, we select the coreset of the Indian Pine hyperspectral image (HSI) by employing *sparse variational inference* [3], and then we train a weighted qSVM on our coreset when using a D-Wave QA. The other version of this work is available in [17].

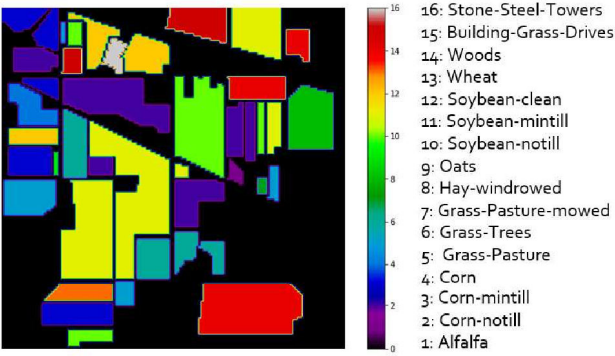

**Fig. 2**. The Indian Pine HSI with 16 classes: {1: Alfalfa, 2: Corn-notill, 3: Corn-mintill, 4: Corn, 5: Grass-Pasture, 6: Grass-Trees, 7: Grass-Pasture-mowed, 8: Hay-windrowed, 9: Oats, 10: Soybean-notill, 11: Soybean-mintill, 12: Soybean-clean, 13: Wheat, 14: Woods, 15: Building-Grass-Drives, 16: Stones-Steel-Towers.

## 2. HSI DATASET

The Indian Pine HSI obtained by AVIRIS sensor has 16 classes, and each class is charachterized by 200 spectral bands shown in Fig. 2 [18]. For simplicity, we used only two-classes of the Indian Pine HSI and charachterized each class by two features by employing Principal Component Analysis (PCA) [13].

## 3. CORESET METHOD

In Bayesian inference, a posterior density $p(\theta|X)$ is expressed by

$$p(\theta|X) = \frac{1}{Z} \exp\left\{\sum_{i=1}^{N} f_i(\theta)\right\} p_0(\theta), \quad (1)$$

where $Z$ is a partition function, $f_i(\theta)$ is a potential function, $p_0(\theta)$ is a prior, $\theta$ parameters, and original data points $\{(X_i, y_i)\}_{i=1}^{N}$ with its labels $y_i$. In practice, a Markov Chain Monte Carlo (MCMC) method is widely used to obtain samples from this posterior due to the intractable $Z$ [19]. To obtain a coreset of the data $\{(X_i, y_i)\}_{i=1}^{N}$, the equation (1) can be re-expressed by

$$p_w = p_w(\theta|X) = \frac{1}{Z(w)} \exp\left\{\sum_{i=1}^{N} w_i f_i(\theta)\right\} p_0(\theta). \quad (2)$$

We denoted the full distribution of the data $\{(X_i, y_i)\}_{i=1}^{N}$ as $p_1 = p_1(\theta|X)$. More importantly, this posterior becomes

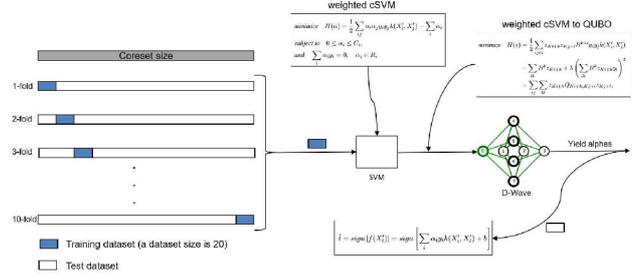

**Fig. 3**. The pictorial representation of our proposed method in this paper.

tractable. Finally, we optimize the below equation via *sparse variational inference* [3]

$$\hat{w} = \min_{w} D_{KL}(p_w || p_1) \quad s.t. \quad w \geq 0, \quad \|w\|_0 \leq M, \quad (3)$$

where $\hat{w}$ is an optimal sparse weight vector, $M$ is a coreset size, and $D_{KL}(p_w || p_1)$ is the Kullback-Leibler (KL) divergence measuring the similarity between (1) and (2).

By optimizing (3), we derive the optimal sparse weight vectors $w$ and the coreset of our dataset such that

$$\{(X_i, y_i)\}_{i=1}^{N} \rightarrow \{(X'_i, w_i, y_i)\}_{i=1}^{M}, \quad w_i \in R_{\geq 0}, \quad (4)$$

where $(X_i, y_i)$ is our original dataset when $(X'_i, w_i, y_i)$ is our coreset.

## 4. WEIGHTED SVM ON THE CORESET

We train a weighted SVM on the coreset expressed by

$$\begin{aligned} \text{minimize} \quad & H(\alpha) = \frac{1}{2} \sum_{ij} \alpha_i \alpha_j y_i y_j k(X'_i, X'_j) - \sum_i \alpha_i \\ \text{subject to} \quad & 0 \leq \alpha_i \leq C_i, \\ \text{and} \quad & \sum_i \alpha_i y_i = 0, \quad \alpha_i \in R, \end{aligned} \quad (5)$$

where $C_i = w_i C$ is a regularization parameter, and $k(\cdot, \cdot)$ is the kernel function of the SVM [19]. This formulation of the SVM is called a kernel-based *weighted* cSVM. The point $X_i$ with $\alpha_i \neq 0$ is called a support vector.

For a test point $X_t$, the decision function is defined by:

$$\hat{t} = sign\left[f(X'_t)\right] = sign\left[\sum_i \alpha_i y_i k(X'_i, X'_t) + b\right], \quad (6)$$

**Table 1**. The coreset of the Indian Pine HSI dataset shown in Fig. 2, and the KL divergence measure.

| Classes | Data size | Coreset Size | KL divergence |
|---------|-----------|--------------|---------------|
| {1, 2}  | 295       | 79           | 0.573         |
| {2, 3}  | 452       | 56           | 0.003         |
| {3, 4}  | 214       | 33           | 0.000         |
| {4, 5}  | 144       | 41           | 0.017         |
| {5, 6}  | 243       | 41           | 0.001         |
| {6, 7}  | 758       | 125          | 0.492         |

**Table 2**. The classification accuracy of the weighted qSVM, and the weighted cSVM on our coreset.

| Classes | Coreset Size | qacc | cacc |
|---------|--------------|------|------|
| {1, 2}  | 79           | 0.96 | 0.96 |
| {2, 3}  | 56           | 0.70 | 0.70 |
| {3, 4}  | 33           | 0.88 | 0.88 |
| {4, 5}  | 41           | 0.78 | 0.78 |
| {5, 6}  | 41           | 0.71 | 0.71 |
| {6, 7}  | 125          | 0.92 | 0.90 |

where $sign(f(X_t')) = 1$ if $f(X_t') > 0$, $sign(f(X_t')) = -1$ if $sign(f(X_t')) < 0$, and $sign(f(X_t')) = 0$ otherwise. The bias $b$ is expressed by following [20]:

$$b = \frac{\sum_i \alpha_i(C_i - \alpha_i)\left[y_i - \sum_j \alpha_j y_j k(X_j', X_i')\right]}{\sum_i \alpha_i(C_i - \alpha_i)}. \quad (7)$$

In addition, we employ a radial basis function (rfb) as a kernel written by

$$rbf(X_i', X_j') = \exp\left\{-\gamma \|X_i' - X_j'\|^2\right\}, \quad (8)$$

where $\gamma > 0$ is a parameter.

## 5. WEIGHTED SVM ON D-WAVE QUANTUM ANNEALER

A D-Wave QA is a quantum annealer with specific topology called *Pegasus*, and its vertices of this topology are occupied by qubits. The D-Wave QA uses a meta-heuristic process to solve only a QUBO problem

$$H(\mathbf{z}) = \sum_{i,j} z_i Q_{ij} z_j, \quad z_i, z_j \in \{0, 1\}, \quad (9)$$

where $z_i, z_j$ are called logical variables, and $Q_{ij}$ includes a bias term $h_i$ and the interaction strength of the logical variables $g_{ij}$ [16]. To solve the weighted cSVM on a D-Wave QA, we transform the weighted cSVM given in (5) to the QUBO problem expressed by (9) by using

$$\alpha_i = \sum_{k=0}^{K-1} B^k z_{Ki+k}, \quad z_{Ki+k} \in \{0, +1\}, \quad (10)$$

where $K$ is the number of binary variables (bits), and $B$ is the base. In the end, the weighted qSVM becomes

$$\begin{aligned} \text{minimize} \quad H(z) &= \frac{1}{2} \sum_{ijkl} z_{Ki+k} z_{Kj+l} B^{k+l} y_i y_j k(X_i', X_j') \\ &\quad - \sum_{ik} B^k z_{Ki+k} + \lambda \left(\sum_{ik} B^k z_{Ki+k} y_i\right)^2 \\ &= \sum_{ij} \sum_{kl} z_{Ki+k} Q_{Ki+k,Kj+l} z_{Kj+l}, \end{aligned} \quad (11)$$

where

$$Q_{Ki+k,Kj+l} = \frac{1}{2} B^{k+l} y_i y_j (k(X_i', X_j') + \lambda) - \delta_{ij} \delta_{kl} B^k. \quad (12)$$

### 5.1. Our experiment

We presented the our KL divergence measure in Table 1. This KL divergence measure implies that the coreset of the Indian Pine HSI is approximately similar to its original dataset. Moreover, we trained the weighted qSVM expressed by (11) and the weighted cSVM written by (5) on this coreset. In addition, we demonstrated our experimental steps of the weighted qSVM in Fig. 3, and we trained the weighted cSVM by using the Python module *scikit-learn*. Finally, we presented our classification results in Table 2, and we noted our quantum accuracy by qacc and our classical accuracy by cacc. This result demonstrates that our coreset is a representative, small subset of its original dataset, and the qSVM is competitive in comparison with its cSVM.

## 6. CONCLUSION

In our paper, we used the coreset of the Indian Pine HSI, and the coreset is a very small and representative weighted subset of its original dataset. We even demonstrated the similarity between our coreset and its original data by computing their KL divergence measure. This similarity implies that we cheaply computed the uncertainty of our original dataset by using its coreset. Finally, we trained the weighted qSVM on our coreset by using a D-Wave QA, and we proved that the weighted qSVM generates classes competitive with the weighted cSVM in Table 2.

## 7. ACKNOWLEDGMENT

The authors gratefully acknowledge the Juelich Supercomputing Centre (https://www.fzjuelich.de/ias/jsc) for funding this project by providing computing time through the Juelich UNified Infrastructure for Quantum computing (JUNIQ) on a D-Wave quantum annealer.